\documentclass[preprint,12pt]{elsarticle}

\usepackage{amssymb}

\journal{Journal of High Energy Density Physics Journal}

\begin{document}

\begin{frontmatter}



\title{Ab Initio Investigation of a Possible Liquid-Liquid Phase Transition in MgSiO$_3$ at Megabar Pressures}


\author{Burkhard Militzer}

\address{Department of Earth and Planetary Science, Department of Astronomy, University of California, Berkeley, CA 94720, USA}

\begin{abstract}
  We perform density functional molecular dynamics simulations of
  liquid and solid MgSiO$_3$ in the pressure range of 120$-$1600 GPa
  and for temperatures up to 20000 K in order to provide new insight
  into the nature of the liquid-liquid phase transition that was
  recently predicted on the basis of decaying laser shock wave
  experiments [Phys. Rev.  Lett. 108 (2012) 065701].  However,
  our simulations did not show any signature of a phase transition in
  the liquid phase. We derive the equation of state for the liquid and
  solid phases and compute the shock Hugoniot curves. We discuss
  different thermodynamic functions and by explore
  alternative interpretations of the experimental findings.
\end{abstract}


\end{frontmatter}


\section{Introduction}
\label{}

The recent work by D. K. Spaulding {\it et al.}~\cite{Spaulding2012}
reported results from decaying laser shock wave experiments which
provided evidence of a liquid-liquid phase transition in MgSiO$_3$ at
megabar pressures. The authors measured a reversal in the shock
velocity and thermal emission and interpreted their findings in terms
of a liquid-liquid phase transition that occurs when the sample
changes from a high density to a low density fluid state during shock
decay. Decaying shock experiments are a new experimental technique
that allows one to map out a whole segment of the Hugoniot curve with
a single shock wave experiment.

First-principles computer simulations have a long tradition of
characterizing materials at extreme
pressures~\cite{MHVTB,KM08,WilsonMilitzer2010} and temperatures~\cite{Mi99,Hu2010,Hu2011} and
of contributing to the interpretation of shock wave experiments~\cite{Mi03,Mi06,Mi09}. The
goal of this particular paper is to perform density functional
molecular dynamics simulations (DFT-MD) of dense liquids~\cite{Mi03}
in order to verify the predictions of a liquid-liquid phase transition
by Spaulding {\it et al.}~\cite{Spaulding2012}. First order
transitions in liquids are unusual but have been seen in experiments
on phosphorus and in simulations of dense hydrogen~\cite{Morales2010}.

\section{Simulation Parameters}

All simulations were performed with the VASP code~\cite{vasp1} with
pseudopotentials of the projector-augmented wave type~\cite{PAW}, a
cut-off for the expansion of the plane wave basis set for the wave
functions of 500~eV, and the PBE exchange-correlation
functional~\cite{PBE}. The Brillioun zone was only sampled with the
$\Gamma$ point to allow for extended and efficient MD simulations. Our
simulations lasted between 1 and 20 ps and used a small time step of
0.2 fs. The electronic states were populated according to the Mermin
functional~\cite{mermin}.

We used an orthorhombic supercell with 60 atoms that we constructed
by tripling the unit cell of the post-perovskite (PPV) structure that
we relaxed at 120 GPa. The system was heated and melted using a
Nos{\'e} thermostat. We then explored the liquid state by scaling the
velocities and changing the density accordingly.

We also performed heat-until-it-melts simulations with 60 atoms
starting a perfect PPV crystal at hydrostatic conditions. We then
gradually increased the temperature in a fixed cell geometry. We also
performed a large number of solid simulations at constant temperature
and different densities.

\begin{figure}[p]
\begin{center}
\includegraphics[width=1.00\textwidth]{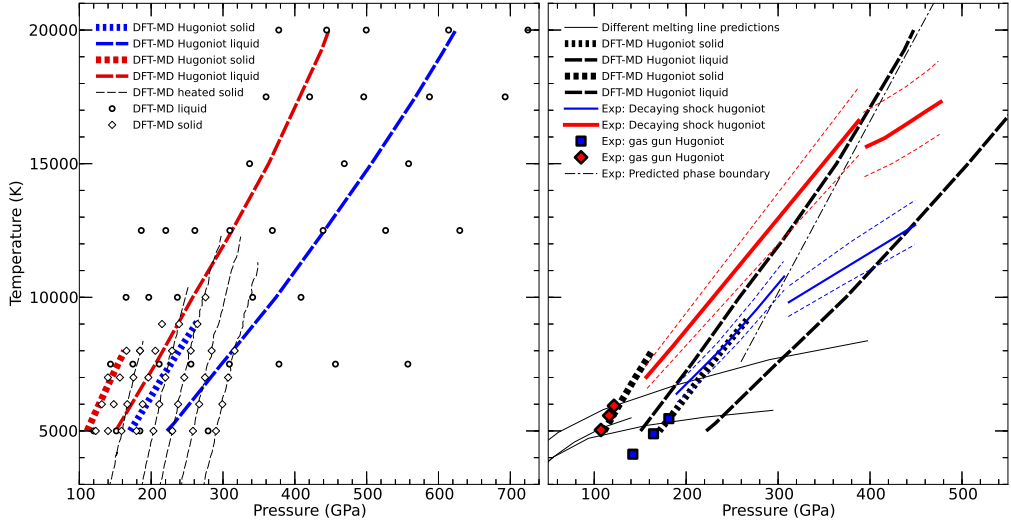}
\end{center}
\caption{ In the left diagram, the open diamond and circles indicate
  the pressure-temperature conditions of our solid and liquid
  MgSiO$_3$ DFT-MD simulations, respectively. The thin black dashed
  lines show the $P$-$T$ paths of the solid portions of our
  heat-until-it-melts simulations. The computed solid and liquid shock
  Hugoniot curves are shown in blue and red color corresponding to
  initial densities of 3.22 and 2.74 $\,$g$\,$cm$^{-3}$ to match those
  of the single crystal (blue) and glass (red) materials used in the
  experiments~\cite{Spaulding2012}.  The Hugoniot curves are repeated
  in black and white on the right for comparison with the decaying
  shock measurements~\cite{Spaulding2012}. From the discontinuities in
  the experimental Hugoniot curves, the phase boundary between a high
  and low density liquid was inferred (dash-dotted line). The computed
  Hugoniot curve for solid MgSiO$_3$ agree well with earlier gas gun
  shock measurements (diamonds)~\cite{GG1,GG2}. Three earlier
  predictions of the melting line are
  included~\cite{melting1,melting2,melting3}. }

\label{PT}
\end{figure}

\section{Results and Discussion}
\begin{figure}[htbl]
\begin{center}
\includegraphics[width=0.90\textwidth]{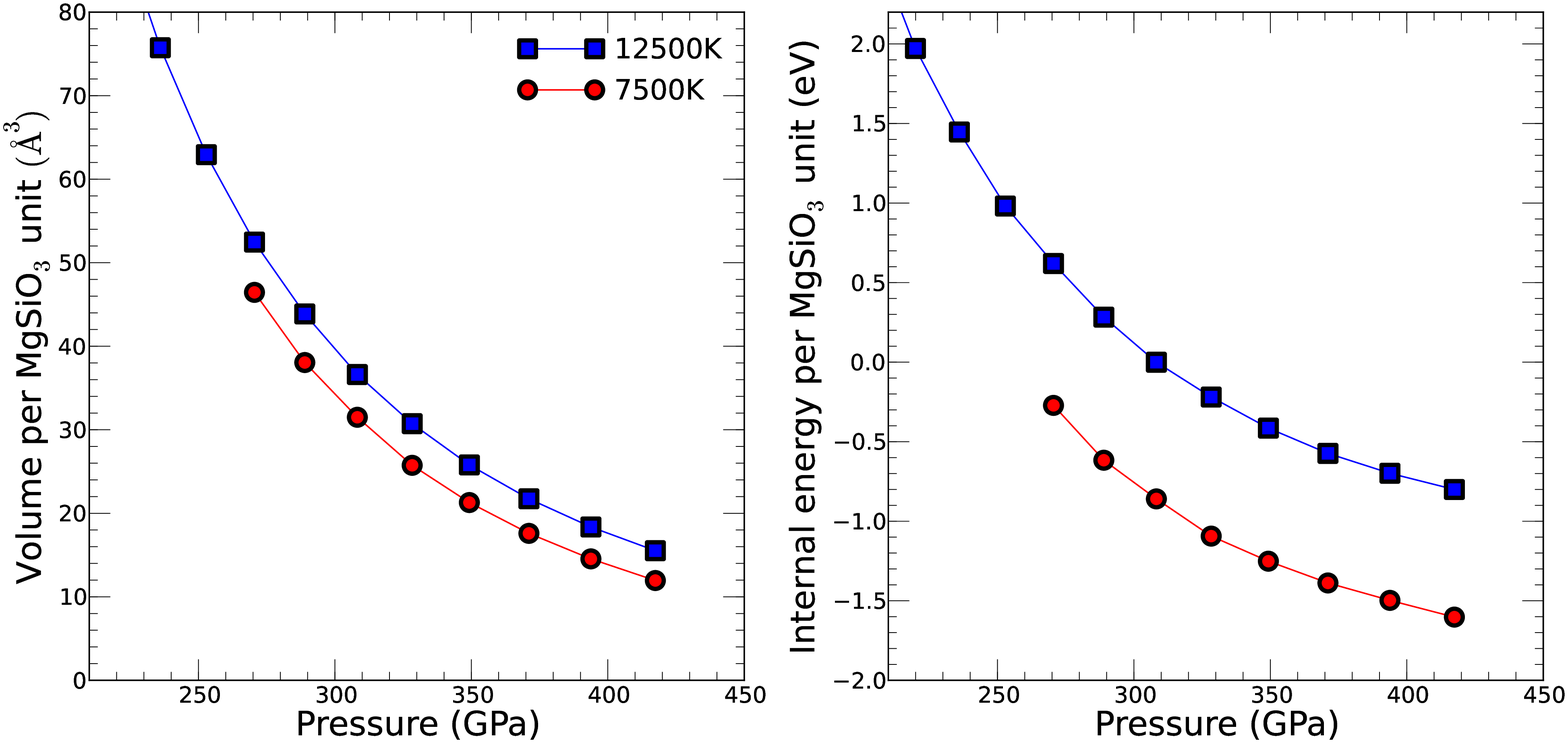}
\end{center}
\caption{Volume and internal energy as function of pressure derived
  from DFT-MD simulations of liquid MgSiO$_3$. Both functions are
  smooth and exhibit no indication of a phase transition.}
\label{VE}
\end{figure}

The $P$-$T$ conditions of our simulations are shown in Fig.~\ref{PT}
and the resulting equation of state (EOS) for the liquid and the
post-perovskite solid phases are given in Tables~\ref{tab1} and
\ref{tab2}.  Initially we focused our work on 7500 K and 12500 K in
order to maximize the likelihood of directly observing a phase
transition in the liquid during the MD simulations.  However, all the
thermodynamic functions, that we analyzed, varied in a perfectly
smooth way as function of temperature and pressure.  Figure~\ref{VE}
shows the volume and the internal energy as function of pressure. The
zero of energy was set equal to the energy of a collection of isolated
atoms. We conclude from this analysis, if there exists a phase
transition in the liquid phase then {\it ab initio} simulations of 60
atoms are unlikely to spontaneously transform into a different phase.

\begin{table}
\centering
{\footnotesize
\begin{tabular}{r r c c r }
$T (K)$ & time (ps) &  $\rho$ (g$\,$cm$^{-3}$) &  $P$ (GPa) &   $E$/FU (eV)\\
\hline
 5000  &  1.47~~~ & 4.7921 & 120.0(5)  &  $-$23.98(7)\\
 5000  &  1.31~~~ & 5.0796 & 151.5(8)  &  $-$22.75(9)\\
 5000  &  1.34~~~ & 5.3905 & 184.3(9)  &  $-$21.91(8)\\
 5000  &  5.02~~~ & 5.7273 & 228.8(5)  &  $-$20.10(4)\\
 5000  &  5.43~~~ & 6.0928 & 279.3(5)  &  $-$17.99(2)\\
\hline
 7500  &  4.90~~~ & 4.7921 & 143.4(5)  &  $-$19.23(7)\\
 7500  &  4.99~~~ & 5.0796 & 174.5(5)  &  $-$17.97(9)\\
 7500  &  4.93~~~ & 5.3905 & 211.2(4)  &  $-$16.66(8)\\
 7500  &  4.66~~~ & 5.7273 & 255.4(5)  &  $-$15.01(8)\\
 7500  &  4.67~~~ & 6.0928 & 308.9(7)  &  $-$13.12(9)\\
 7500  &  5.94~~~ & 6.4900 & 377.9(6)  &  $-$10.31(7)\\
 7500  &  5.60~~~ & 6.9225 & 456.6(8)  &  $-$7.40(9)\\
 7500  &  5.78~~~ & 7.3944 & 557.4(7)  &  $-$3.26(9)\\
\hline
10000  &  1.17~~~ & 4.7921 & 165.0(8)  &  $-$14.41(12)\\
10000  &  1.24~~~ & 5.0796 & 196.6(1.1)  &  $-$13.24(19)\\
10000  &  1.14~~~ & 5.3905 & 236.5(9)  &  $-$11.94(11)\\
10000  &  4.94~~~ & 6.0928 & 341.3(7)  &  $-$7.63(6)\\
10000  &  5.30~~~ & 6.4900 & 408.6(9)  &  $-$5.10(13)\\
\hline
12500  & 22.94~~~ & 4.7921 & 186.2(3)  &  $-$9.61(4)\\
12500  & 17.24~~~ & 5.0796 & 220.3(4)  &  $-$8.38(6)\\
12500  & 12.72~~~ & 5.3905 & 260.8(3)  &  $-$6.88(7)\\
12500  & 17.99~~~ & 5.7273 & 309.5(3)  &  $-$4.98(5)\\
12500  & 15.74~~~ & 6.0928 & 368.7(3)  &  $-$2.63(6)\\
12500  &  3.52~~~ & 6.4900 & 439.4(1.0)  &  $-$0.01(13)\\
12500  &  3.63~~~ & 6.9225 & 526.4(1.2)  &  3.39(18)\\
12500  &  4.36~~~ & 7.3944 & 629.7(1.0)  &  7.44(14)\\
12500  &  4.77~~~ & 7.9101 & 755.2(8)  &  11.78(6)\\
12500  &  4.72~~~ & 8.4749 & 908.8(1.2)  &  17.36(17)\\
12500  &  4.80~~~ & 9.0948 & 1094.4(1.8)  &  23.66(17)\\
12500  &  5.02~~~ & 9.7766 & 1326.5(1.7)  &  31.8(2)\\
12500  & 18.83~~~ & 10.5283 & 1604.8(1.1)  &  41.18(14)\\
\hline
15000  &  3.34~~~ & 5.7273 & 337.0(1.0)  &  0.11(16)\\
15000  &  3.75~~~ & 6.4900 & 468.9(8)  &  5.01(10)\\
15000  &  3.88~~~ & 6.9225 & 558.4(1.1)  &  8.56(14)\\
\hline
17500  &  1.49~~~ & 5.7273 & 359.9(1.7)  &  4.48(20)\\
17500  &  3.30~~~ & 6.0928 & 420.6(8)  &  6.64(11)\\
17500  &  2.18~~~ & 6.4900 & 495.9(1.0)  &  9.78(14)\\
17500  &  3.65~~~ & 6.9225 & 587.5(1.1)  &  13.42(13)\\
17500  &  3.69~~~ & 7.3944 & 692.8(1.1)  &  16.96(14)\\
\hline
20000  &  1.41~~~ & 5.7273 & 377.6(1.4)  &  8.3(2)\\
20000  &  2.70~~~ & 6.0928 & 444.3(6)  &  11.10(11)\\
20000  &  4.62~~~ & 6.4900 & 499.4(6)  &  9.21(11)\\
20000  &  3.38~~~ & 6.9225 & 614.1(1.2)  &  17.96(18)\\
20000  &  3.56~~~ & 7.3944 & 724.5(1.0)  &  22.06(14)\\
\hline
        \end{tabular}
} 
\caption{Temperature, MD simulation time, density,
  pressure and internal energy from our DFT$-$MD simulations of liquid MgSiO$_3$. The
  1$\sigma$ uncertainties of the trailing digits are given in brackets.}
\label{tab1}
\end{table}
                                       
\begin{table}
\centering
{\footnotesize
\begin{tabular}{r r c c r }
$T (K)$ & time (ps) &  $\rho$ (g$\,$cm$^{-3}$) &  $P$ (GPa) &   $E$/FU (eV)\\
\hline
 5000  &  4.23~~~ & 5.0796 & 122.86(16)  &  $-$26.135(11)\\
 5000  &  4.33~~~ & 5.2320 & 139.94(19)  &  $-$25.591(12)\\
 5000  &  4.59~~~ & 5.3905 & 158.71(10)  &  $-$24.927(5)\\
 5000  &  3.85~~~ & 5.5555 & 179.59(14)  &  $-$24.196(9)\\
 5000  &  5.00~~~ & 5.7273 & 203.19(16)  &  $-$23.339(9)\\
 5000  &  4.74~~~ & 5.9063 & 229.46(12)  &  $-$22.380(8)\\
 5000  &  5.00~~~ & 6.0928 & 258.19(18)  &  $-$21.293(5)\\
 5000  &  5.00~~~ & 6.2872 & 290.29(14)  &  $-$20.068(7)\\
\hline
 6000  &  4.18~~~ & 5.0796 & 131.24(15)  &  $-$24.737(13)\\
 6000  &  4.29~~~ & 5.2320 & 147.89(14)  &  $-$24.217(7)\\
 6000  &  4.50~~~ & 5.3905 & 166.75(17)  &  $-$23.584(9)\\
 6000  &  3.83~~~ & 5.5555 & 188.2(2)  &  $-$22.825(14)\\
 6000  &  4.75~~~ & 5.7273 & 211.29(14)  &  $-$21.998(8)\\
 6000  &  4.73~~~ & 5.9063 & 237.50(15)  &  $-$21.049(8)\\
 6000  &  5.00~~~ & 6.0928 & 266.9(2)  &  $-$19.979(6)\\
 6000  &  5.00~~~ & 6.2872 & 299.11(14)  &  $-$18.748(5)\\
\hline
 7000  &  4.11~~~ & 5.0796 & 139.7(2)  &  $-$23.25(3)\\
 7000  &  4.92~~~ & 5.2320 & 156.41(12)  &  $-$22.768(17)\\
 7000  &  4.86~~~ & 5.3905 & 175.31(15)  &  $-$22.170(14)\\
 7000  &  3.81~~~ & 5.5555 & 196.05(13)  &  $-$21.410(17)\\
 7000  &  4.73~~~ & 5.7273 & 219.72(14)  &  $-$20.632(10)\\
 7000  &  4.64~~~ & 5.9063 & 246.1(2)  &  $-$19.672(9)\\
 7000  &  5.00~~~ & 6.0928 & 274.90(12)  &  $-$18.611(7)\\
 7000  &  5.00~~~ & 6.2872 & 307.1(2)  &  $-$17.397(12)\\
\hline
 8000  &  4.87~~~ & 5.2320 & 165.6(2)  &  $-$21.17(3)\\
 8000  &  4.39~~~ & 5.3905 & 184.5(3)  &  $-$20.58(4)\\
 8000  &  4.39~~~ & 5.3905 & 184.5(3)  &  $-$20.58(4)\\
 8000  &  3.75~~~ & 5.5555 & 205.8(3)  &  $-$19.90(3)\\
 8000  &  5.00~~~ & 5.7273 & 229.08(20)  &  $-$19.122(18)\\
 8000  &  4.62~~~ & 5.9063 & 255.10(19)  &  $-$18.204(15)\\
 8000  &  5.00~~~ & 6.0928 & 284.2(2)  &  $-$17.163(17)\\
 8000  &  5.00~~~ & 6.2872 & 316.3(2)  &  $-$15.964(12)\\
\hline
 9000  &  3.62~~~ & 5.5555 & 215.1(3)  &  $-$18.27(2)\\
 9000  &  5.00~~~ & 5.7273 & 238.9(3)  &  $-$17.50(2)\\
 9000  &  4.26~~~ & 5.9063 & 264.4(3)  &  $-$16.63(2)\\
\hline
10000  &  4.53~~~ & 5.9063 & 275.6(3)  &  $-$14.84(3)\\
\hline
        \end{tabular}
} 
\caption{Temperature, MD simulation time, density,
  pressure and internal energy from our DFT-MD simulations of crystalline post-perovskite MgSiO$_3$. The
  1$\sigma$ uncertainties of the trailing digits are given in brackets.}
\label{tab2}
\end{table}

We computed the shock Hugoniot curve from the usual
relation~\cite{Ze66},
\begin{equation}
H = E-E_0+\frac{1}{2}(V-V_0)(P+P_0)=0,
\label{hug}
\end{equation}
where the initial volume, $V_0$=51.77 \AA$^3$, and initial internal
energy, $E_0$ = $-$35.914 eV per MgSiO$_3$ formula unit (FU), were
taken from a DFT calculation of enstatite at 3.22 g$\,$cm$^{-3}$ that
we performed. We used the same $E_0$ to derive the Hugoniot curves for
a glass sample with an initial density of 2.74 g$\,$cm$^{-3}$.  $P_0$
was assumed to be zero because $P_0 \ll P$ is satisfied for all final
states.

The resulting DFT-MD Hugoniot curves for liquid and crystalline
MgSiO$_3$ are shown in Fig.~\ref{PT}. Our results for the solid phase
agree well with the earlier shock measurements using a gas
gun~\cite{GG1,GG2}. However, the agreement with the decaying laser
shock results is not satisfactory. A portion of our computed Hugoniot
curve for the solid at $\rho_0=3.22$ g$\,$cm$^{-3}$ overlaps with the
measurements in the temperature range from 7000 to 9000 K.  Also at
450 GPa and 12500 K, the calculated Hugoniot curve for the liquid
agrees with the measurements but the slope of the theoretical Hugoniot
curve is different and it shows no sign of a phase transition. The
agreement for an initial density of 2.74 g$\,$cm$^{-3}$ is again not
favorable. The theoretical Hugoniot the liquid passes right through
the $T$-$P$ conditions where a phase transformation is predicted based
on the experimental data but DFT-MD simulations do not exhibit one.

\begin{figure}[htbl]
\begin{center}
\includegraphics[width=0.90\textwidth]{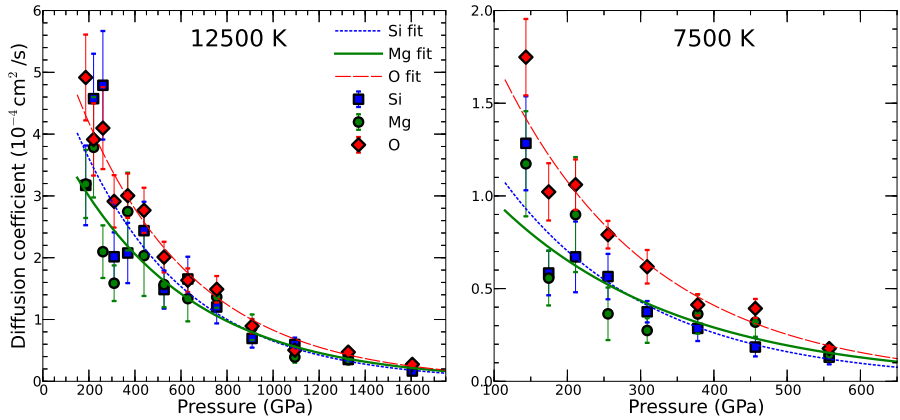}
\end{center}
\caption{Diffusion coefficients of Mg, Si, and O ions as function of
  pressure for 7500 and 12500 K inferred from DFT-MD simulations. The
  lines show fits to an exponential function of pressure for each
  species.}
\label{D}
\end{figure}

Next we investigated the possibility of the existence of a superionic
phase between the solid and the liquid phases where some ions remain
stationary while others diffuse throughout the material like in a
liquid~\cite{cavazzoni}. In Fig.~\ref{D}, we plot the diffusion
coefficients as a function of pressure. At 12500 K, all ions diffuse
at the same rate approximately.  At 7500 K, the diffusion is slower as
one would expect in the fluid near the melting line. The oxygen ions
diffuse a bit faster than the magnesium and silicon ions but there is
no evidence for a superionic state.

\begin{figure}[htbl]
\begin{center}
\includegraphics[width=0.80\textwidth]{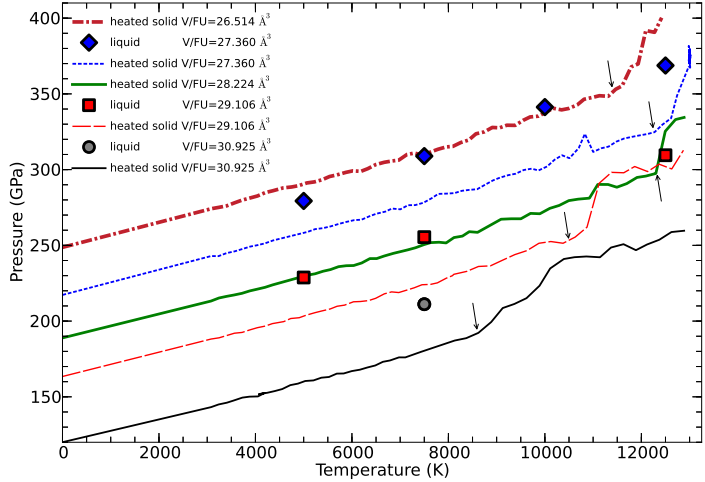}
\end{center}
\caption{The pressure-temperature behavior during heat-until-it-melts
  simulations (lines) is compared with isothermal liquid simulations
  (symbols). The small arrows indicate the beginning of the melting
  transformation in the MD simulation. Liquid samples are found to
  always exhibit a higher pressure than solid ones at the same
  density.}
\label{heating}
\end{figure}

We also performed heat-until-it-melts simulations of solid samples
(Fig.~\ref{PT}) that we isochorically heated at a rate of 1000 K per
picosecond. This approach was used to predict the superionic state of
water~\cite{cavazzoni} but our sample always went into a fully fluid
state after a substantial amount of superheating. Figure~\ref{heating}
shows that the liquid always exhibits a higher pressure than the solid
when compared for the same density and temperature. This means, for
given temperature and pressure, the solid is always denser than the
liquid. Therefore the melting line of PPV MgSiO$_3$ is expected to
have a positive Clapeyron slope~\cite{Tateno2009}.

We performed our simulations with only 60 atoms and all results may
need to be confirmed with larger simulations that possibly also use
more k-points. Our longest simulations ran for 20 ps, which is fairly
long for {\it ab initio} calculations. If a new liquid phase emerges,
one would expect to observe a spontaneous phase transformation, as
long as it does not involve any large-scale rearrangements of atoms.
If liquid MgSiO$_3$ phase-separates into a MgO-dominated and a
SiO$_2$-rich fluid, as was predicted for the solid at approximately
1000 GPa~\cite{Umemoto2006}, we would most likely not be able to
observe this process in our simulations. One would instead need to
perform more expensive Gibbs free energy
calculations~\cite{Alfe2000,Morales2009,WilsonMilitzer2010,WilsonMilitzer2012,WilsonMilitzer2012b}
but such a complex effort goes beyond this investigation.

Furthermore, if there existed a new, unknown phase in the MgSiO$_3$
phase diagram, it does not have to be a liquid. One could imagine a
new stable solid phase that introduces a solid-solid-liquid triple
point and may lead to an increase in the slope of the melting line.
At zero temperature, existence of a post-post-perovskite phase has
been intensely studied with {\it ab initio} random structure search
algorithms and no such phase has been found but a new,
entropy-stabilized phase may still exist at high temperature. In
principle, this solid-solid-liquid triple point could also be between
perovskite (PV), PPV, and liquid phases. This point has not yet been
determined neither with experimental nor with theoretical means but
PV-to-PPV transition pressure is known to increases with temperature.

One may also ask whether there exists an alternative interpretation
for the experimental observations. Our first recommendation would be
to repeat those measurements with steady shock wave experiments in
order to verify the discontinuities on the principal Hugoniot curve of
MgSiO$_3$. This may require a series of shock experiments with
relatively small error bars but it would be an important confirmation.

It is also possible that observed shock velocity reversal is related
to the melting transition of MgSiO$_3$. Figure~\ref{PT} shows that
most of the experimental results fall in between the computed Hugoniot
curves for the solid and the liquid. However, the temperatures are too
high for a solid phase to be thermodynamically stable. Phase
transition temperatures of 10$\,$000$\,$ and 16$\,$000$\,$K have been
predicted based on the measurements of single-crystal and glass
material~\cite{Spaulding2012}, respectively. Nevertheless it is
interesting to discuss the behavior of decaying shock experiments in
the presence of a melting transition. Some material behind the shock
front may freeze during the shock decay. This could introduce a
secondary wave and thereby affect the behavior of the shock front.

\begin{figure}[htbl]
\begin{center}
\includegraphics[width=0.40\textwidth]{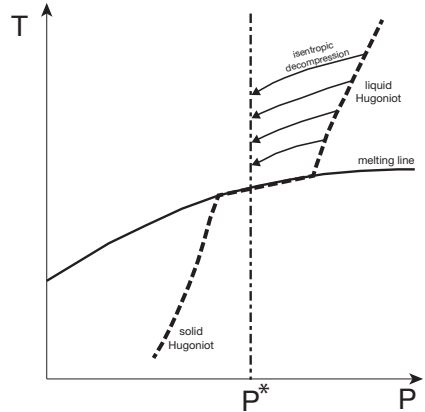}
\end{center}
\caption{Temperature-pressure diagram with melting lines and shock
  Hugoniot curves. The arrows indicate the adiabatic and isentropic
  expansion of material during decaying shocks. The graph shows a
  material with normal melting behavior with a positive Clapeyron
  slope, $dT_m/dP>0$.}
\label{diagram}
\end{figure}

At the beginning of a decaying shock experiments, the sample material
is compressed to a state of high pressure and high temperature on the
principal Hugoniot curve. As the shock decays, new material does not
reach as high pressures and temperatures but stays on the
Hugoniot curve, which allows one to map out the whole Hugoniot curve
with just one shock measurement. The question is what happens to the
material that had compressed to a higher $P$-$T$ state earlier. One
may assume that the whole region behind the shock front will
equilibrate to a new pressure, which we labeled $P^*$ in
Fig~\ref{diagram}. Any parcel of material that was shocked to a high
$P$-$T$ state will adiabatically expand~\cite{Ze66} to reach $P^*$ and
slow down to travel at the new and reduced particle velocity. Since
this expansion is gradual and not associated with any shock, one
typically also assumes that this expansion is isentropic~\cite{Ze66} (arrows in
Fig.~\ref{diagram}). While one assumes a new pressure, $P^*$, is
established, the entire sample behind the shock front is not expected
to reach thermal equilibrium during the experiment. This leads to the
situation where hot material at a lower density is pushing on colder
material at a higher density. This could lead to a fluid dynamic
instability of Rayleigh-Taylor type if the density contrast becomes
too high and the time scale is long enough for such an instability to
develop.

It is important to note that Hugoniot curves are steeper
in $P$-$T$ space than isentropes, $\left. \frac{\partial T}{\partial
    P}\right|_{\rm Hug} > \left.\frac{\partial T}{\partial
    P}\right|_{S}$. The shock front on the Hugoniot curve would always enter
any new thermodynamic phase, that may exist at lower temperature,
before the hotter material behind the shock front reaches it.

\begin{figure}[htbl]
\begin{center}
\includegraphics[width=1.00\textwidth]{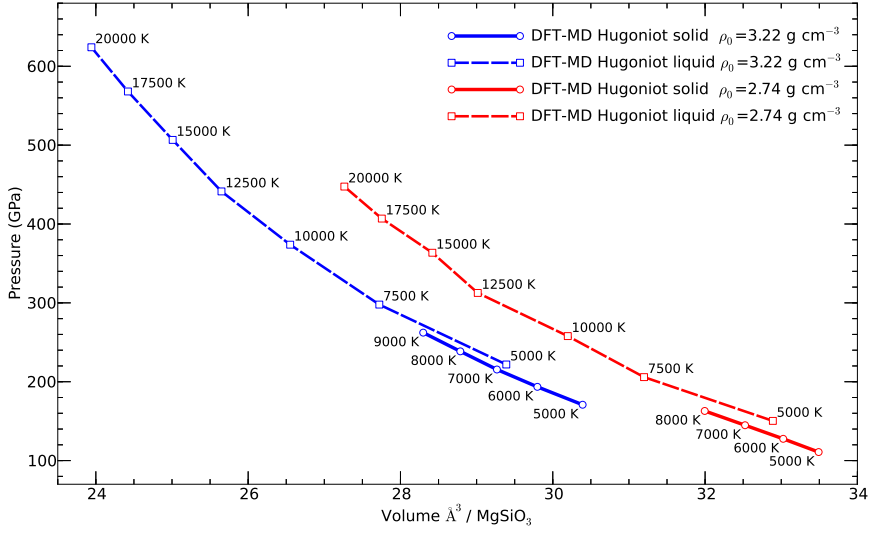}
\end{center}
\caption{$P$-$V$ diagram of the solid and liquid segments of the shock Hugoniot curves for
  two initial densities. The labels indicate the calculated shock
  temperatures.}
\label{PV}
\end{figure}

\begin{figure}[htbl]
\begin{center}
\includegraphics[width=1.00\textwidth]{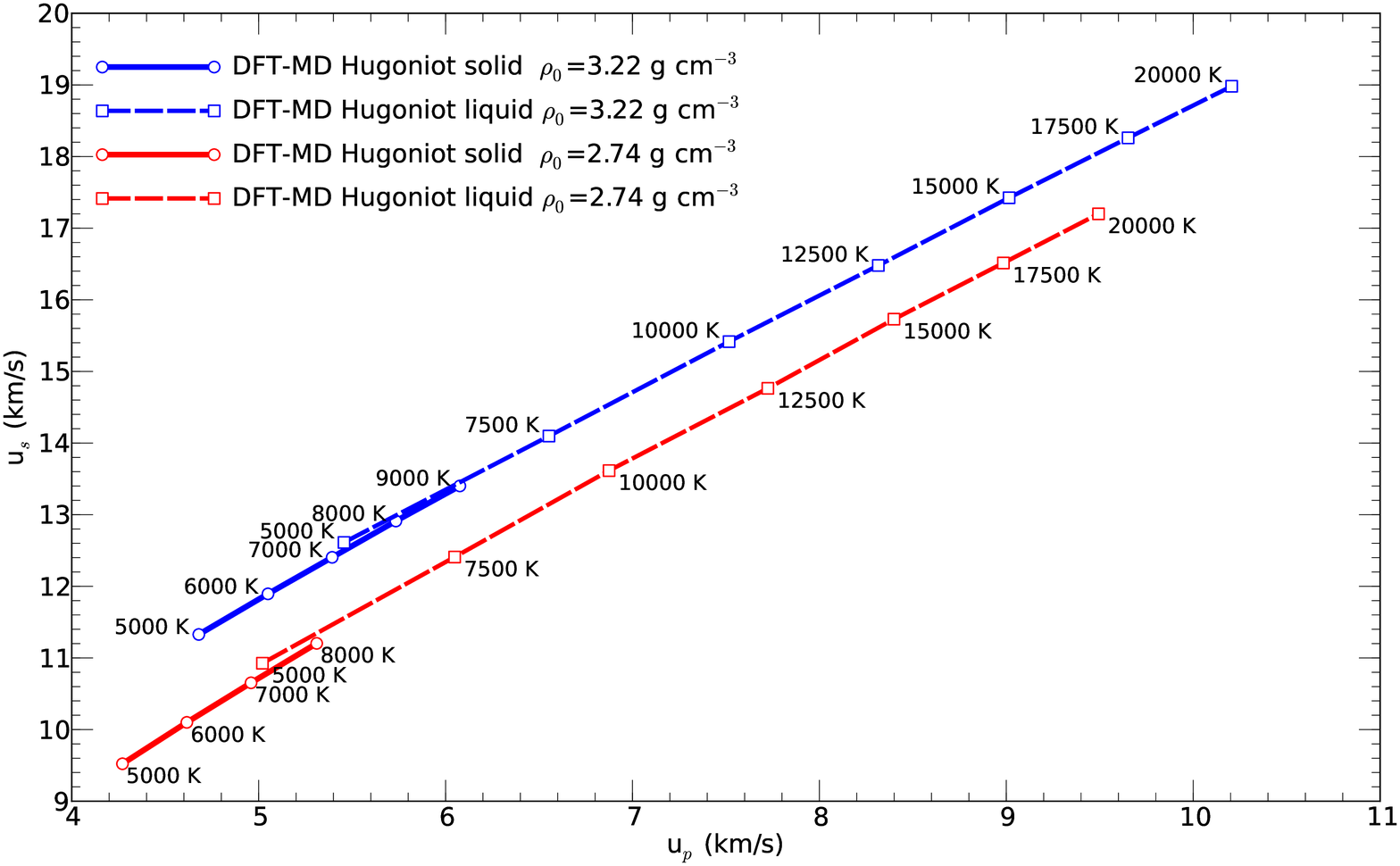}
\end{center}
\caption{Shock velocity versus particle velocity for the Hugoniot curves from Fig.~\ref{PV}.}
\label{usup}
\end{figure}

However, the melting transition may have an effect on the
shock propagation at much higher temperature.  Zeldovich and
Raizer~\cite{Ze66} showed that a shock will split into two waves if
the shock Hugoniot curves intersects with the Rayleigh line that
linearly connects the initial and final states in a $P$-$V$ diagram.
If the melting transition would introduce such a splitting, it would
affect the shock propagation at much higher temperatures and
pressures. We plotted our liquid and solid Hugoniot curves the $P$-$V$
and $u_s$-$u_p$ diagrams in Figures~\ref{PV} and \ref{usup}. For each
of the Hugoniot curves, let us assume that material melts at a
temperature, $T_m$, somewhere in the range spanned by the calculation
in Refs.~\cite{melting1,melting2,melting3} (see Fig.~\ref{PT}). If
there occurred a sudden and complete phase change from liquid to solid
during a decaying shock experiments at $T_m$, it would lead to a
abrupt drop in pressure, density, shock and particle velocities
because intermediate values cannot be realized by either phase while
satisfying Eq.~(\ref{hug}). 

From Fig.~\ref{PV}, we conclude that the Hugoniot curves do not
intersect with the Rayleigh line but the material must assume a mixed
state because the particle velocity decreases gradually in a decaying
shock wave experiment. Following the arguments that Greeff {\it et
  al.}~\cite{Greeff2001} applied to the $\alpha$-to-$\omega$
transition in solid titanium, the properties of a solid-liquid
MgSiO$_3$ mixture with the solid fraction, $\lambda$, can be derived
from
\begin{eqnarray}
E &=& (1-\lambda)E_l(P,T) + \lambda E_s(P,T)\\
V &=& (1-\lambda)V_l(P,T) + \lambda V_s(P,T)
\end{eqnarray}
for a given pair of $T$ and $P$ on the melting line. Under the
assumption of thermodynamic equilibrium, $\lambda$ can derived from
Eq.~(\ref{hug}) for given pressure or particle velocity. For high shock
velocities, the final state of the material is fully molten. As the
shock decays, the solid fraction gradually increases until the shock
is too weak to introduce any melting. 

We now want to address the question what happens as the parcels of
solid-liquid mixtures expand adiabatically. The solid fraction, $\lambda$,
must change to keep the entropy of the mixture,
\begin{eqnarray}
S &=& (1-\lambda)S_l(P,T) + \lambda S_s(P,T)\;\;,
\label{S}
\end{eqnarray}
constant as the pressure decreases. $T$ adjusts to the melting
temperature for given pressure. If the solid fraction, $\lambda$,
gradually increases during the expansion, the particle velocity
directly behind the shock front would decrease. This is not unusual
but the opposite case is much more interesting. If the liquid fraction
increases during the shock decay, $\left. \frac{\partial
    \lambda}{\partial P}\right|_{T_m}>0$, it would imply an increase
in particle velocity. As a result the shock wave would split in two
waves, which would make the analysis of the VISAR signals much more
difficult.

We can determine the thermodynamic parameters for such a shock wave
splitting. During the expansion, the total entropy stays constant,
$\left. \frac{\partial S}{\partial P}\right|_{T_m}=0$. Eq.~(\ref{S})
implies,
\begin{eqnarray}
\left. \frac{\partial \lambda}{\partial P}\right|_{T_m} (S_l-S_s) 
= (1-\lambda) \left. \frac{\partial S_l}{\partial P}\right|_{T_m} 
+ \lambda     \left. \frac{\partial S_s}{\partial P}\right|_{T_m} 
\label{dS}
\end{eqnarray}
Since the entropy of the liquid is always greater than that of the
solid, a positive right term in Eq.~(\ref{dS}) would imply that the
shock wave splits into two separate waves as the decay shock
experiment enter the regime of solid-liquid coexistence. Without a
detailed calculations of the entropies in both phases, it is not
possible to determine whether this occurs in MgSiO$_3$ because the
Clapeyron slope enters into the calculation. However, the entropy is
accessible with thermodynamic integration techniques and equation of
state models~\cite{melting3}. So our hypothesis of shock wave
splitting can be tested with experimental and theoretical techniques.

Summarizing we can say that whether a Rayleigh-Taylor instability
exists, a new solid phase appears, the shock wave splits into two
waves, or indeed a liquid-liquid phase transition is the reason for
the observed shock velocity reversal remains to be determined with
future experiments. Concluding our theoretical investigation, we can
report that our DFT-MD simulations to not support the hypothesis of a
liquid-liquid phase transition in MgSiO$_3$. We were not able to give
an obvious alternative interpretation of the experimental findings but
we do, however, suggest the measurements be repeated with steady shock
waves. The experimental findings are very interesting nevertheless and
will lead to a better understanding of magnesiosilicates at high
temperature.

\appendix
\section*{Acknowledgments}

The work was supported by NSF and NASA. We thank D. K. Spaulding and
R. Jeanloz for disucssions and comments on this manuscript.












\end{document}